\documentclass[letter]{ptptex}
\usepackage{graphicx}
\usepackage{wrapft,amsmath,amssymb}
      
\def\be{\begin{equation}}
\def\ee{\end{equation}}
\def\bea{\begin{eqnarray}}
\def\eea{\end{eqnarray}}
\def\ba{\begin{array}}
\def\ea{\end{array}}
\def\<{\left\langle}
\def\>{\right\rangle}
\def\({\left(}
\def\){\right)}
\def\e{{e}}

\def\openone{\mathbf{1}}

\notypesetlogo  
\title{Level Spacings of Parametric Chiral Random Matrices\\
and Two-Color QCD with Twisted Boundary Condition}

\author{Shinsuke M. \textsc{Nishigaki}}

\inst{
Graduate School of Science and Engineering, Shimane University, \\
Matsue 690-8504, Japan}

\recdate{August 20, 2012; Revised November 22, 2012}

\abst{We evaluate the level spacing and smallest eigenvalue distributions
of chiral random matrix ensembles transiting from symplectic or 
orthogonal to unitary symmetry classes with a crossover parameter $\rho$.
As expected from the effective $\sigma$ model description,
these results can be fitted perfectly to the fundamental or adjoint 
staggered Dirac spectrum of $SU$(2) quenched lattice gauge theory 
under the imaginary chemical potential (twisting) $\mu$.
The linear dependence of the parameter $\rho$ on $\mu$ determines 
the pion decay constant $F$ as its proportionality constant. }

\PTPindex{100, 138, 160, 169, 183, 350}  

\begin{document}
\maketitle

\noindent
{\it Introduction}\hspace*{7mm}
Two-color QCD has served as a strategic testing ground for realistic chromodynamics,
as well as a tractable model interesting in its own right,
owing to the absence of the sign problem at finite density and with pairs of degenerated flavors\cite{nak}.
This and other gauge theories with quarks in a (pseudo)real representation
exhibit exotic global symmetry breaking\cite{peskin}:
quarks and charge-conjugated antiquarks form a multiplet of the extended flavor group, 
which breaks spontaneously to the extended vector subgroup\cite{vw}.
Accordingly, the effect of the chemical potential that distinguishes quarks from antiquarks and
breaks this extended symmetry 
is incorporated into the low-energy chiral Lagrangian\cite{ls}
through the flavor-covariant derivative\cite{kogut,dn}.
Because the unconventional global symmetries originate from the presence of
antiunitary symmetries of Dirac operators,
the symmetry-violating chemical potential $\mu$
in these QCD-like theories also manifests itself in the statistical properties of Dirac spectra\cite{sv}.
One can indeed predict the fluctuation of Dirac eigenvalues
that permeate into the complex plane
from the zero-momentum part of chiral Lagrangians, which
in turn is equivalent to the chiral Gaussian orthogonal or symplectic ensemble (chGOE, chGSE)\cite{ver}
extended to its non-Hermitian counterpart
through the introduction of a schematic density component coupled to (real) $\mu$.\cite{hjv,ake}.

Antiunitary symmetries
of $SU$(2) Dirac operators could, however,
also be violated by any {\it Hermitian} term in a complex representation,
most simply by the $U$(1) gauge field either fluctuating or fixed as a background,
the latter being gauge-equivalent to
the twisted boundary condition or imaginary chemical potential
\cite{jnpz,svd,mt}.
In this case, Dirac eigenvalues remain real after the inclusion of symmetry violation
and their statistical behavior exhibits crossover
instead of permeation to the complex plane.
Since these two cases are essentially identical
in the chiral Lagrangian description save for the sign of $\mu^2$,
studies of Dirac spectra from both sides should reveal the validity of
analytic continuation in $\mu$, as is required for three-color QCD.

Spectral crossover between universality classes of Hermitian random matrix (RM) ensembles
\cite{dys2,mp,meh}, namely from GOE or GSE to GUE,
has been extensively studied for disordered and chaotic Hamiltonians 
whose time-reversal invariance is slightly broken by a magnetic field.
The transitional behavior of spectral fluctuations
is by itself universal in the sense that the local spectral fluctuation depends
only on a single parameter $\rho$ defined below.
The reason for this universality is traced back to
the nonlinear $\sigma$ model (Eq.~(\ref{Zchiral}) below) governing the spectral statistics,
which can be derived either
by conventional disorder averaging \cite{aie} or
by summation over encountered periodic orbits of chaotic systems \cite{snmb},
without any reference to the details of dynamics.
Thus, one can rely on the simplest model that yields the identical $\sigma$ model,
i.e.,\ parametric RMs, for actual computation.

On the other hand, chiral or superconducting variants of universality crossover
have been relatively less explored.
Damgaard and collaborators\cite{dam1,dam2}
have achieved a significant breakthrough by analytically computing
correlation functions and individual small eigenvalue distributions
for the spectral crossover within the chiral Gaussian unitary ensemble (chGUE) class. 
They presented convincing numerical evidence that
the Dirac spectrum of three-color QCD at an imaginary isospin chemical potential
indeed exhibits crossover as predicted by chiral RMs.
On the other hand,
although analytic results for microscopic spectral correlation functions
for the crossover from chGOE or chGSE to chGUE 
have been known for some time \cite{nag,kt},
they have lacked physical application.
To the best of our knowledge, the only physical example of crossover involving 
{\em different} Hermitian chiral universality classes is the CI-C transition for the
super/normal/superconducting hybrid interface in a magnetic field \cite{koz}.
In this paper, we provide novel applications of crossover between chiral Hermitian universality classes
from lattice gauge theory.

\vspace{1mm}
\noindent
{\it Parametric chiral random matrices}\hspace*{7mm}
Consider an ensemble of $N\times N$ Hermitian complex (quaternion) matrices $H=H_{\rm S}+i\alpha H_{\rm A}$,
with $H_{\rm S}$ real symmetric (quaternion self-dual) 
and $H_{\rm A}$ real antisymmetric (quaternion anti-self-dual),
distributed according to Gaussian measures of variance $\sigma^2$.
By setting ${N}/{2}\times {N}/{2}$ block-diagonal parts of $H$ to zero,
the matrix takes the form (${}^{T}$ and ${}^{D}$ stand for transpose and quaternion-dual, respectively)
\[
H=
\left[
\ba{cc}
0&H_1+i\alpha H_2\\
(H_1-i\alpha H_2)^{T,D} &0
\ea
\right],\ \ H_1, H_2 : \frac{N}{2}\times \frac{N}{2}\ \mbox{(quaternion-) real matrix}.
\]
Depending on the parameter $\alpha$,
this parametric (also called dynamical or Brownian motion \cite{dys2}) chiral RM ensemble
is
used to interpolate between two limiting cases,
chGOE (chGSE) at $\alpha=0$ and chGUE at $\alpha=1$.
Since nonzero eigenvalues of $H$ occur in pairs of equal magnitude and opposite signs,
it suffices to retain nonnegative eigenvalues only.
The correlation function of $n$ eigenvalues $\{\lambda_i\}$ of $H$ in the vicinity of the origin 
(where the mean level spacing is $\varDelta(0)$)
is expressed,
in the limit $N\to\infty,\ \alpha\to 0$ and $\rho\equiv {\alpha \sigma}/{\varDelta(0)}$ fixed,
as a determinant with $S, D, I$ given by\cite{nag,kt}
{\small
\bea
&&
R_{n}(x_1,\ldots,x_n)=
\bigl(\det \left[K(x_i, x_j)\right]_{i,j=1}^n\bigr)^{1/2}~,
\ \ 
K(x, y)=\left[
\ba{cc}
S(x,y) & I(x, y)\\
D(x,y) & S(y, x)
\ea
\right],
\label{Pfaffian}\\
&&S(x,y)=\pi  \sqrt{x y} \left\{
\frac{x J_1(\pi  x) J_0(\pi  y)-J_0(\pi  x)y J_1(\pi  y)}{x^2-y^2}
+\frac{J_0(\pi y)}{2}\!\!\int_\pi^\infty \!\!\!dv\,\e^{-\rho^2 (v^2-\pi^2)}J_0(v x)\right\}
,\nonumber\\
&&D(x,y)=
-\frac{\sqrt{x y} }{2}  \int_0^\pi dv\, v^2 \,\e^{2 \rho^2 v^2} 
\left\{x J_1(v x) J_0(v y)- J_0(v x)y J_1(v y)\right\}     
,\nonumber\\
&&I(x,y)=
\frac{\sqrt{xy}}{2}  \int_\pi^\infty dv\,v  \int_1^\infty du\,\e^{-\rho^2 v^2(1+u^2)}
\left\{J_0(v u x) J_0(v y)- J_0(v x) J_0(v u y)\right\}
\label{chGOEchGUE}
\end{eqnarray}}
for chGOE-chGUE crossover, and
{\small
\begin{eqnarray}
&&S(x,y)=\pi  \sqrt{x y} \left\{
\frac{x J_1(\pi  x) J_0(\pi  y)-J_0(\pi  x)y J_1(\pi  y)}{x^2-y^2}
-\frac{J_0(\pi x)}{2}\!\!\int_0^\pi \!\!\!dv\,\e^{\rho^2 (v^2-\pi^2)}J_0(v y)\right\}
,\nonumber\\
&&D(x,y)=
\frac{\sqrt{xy} }{2} \int_0^\pi dv\,v  \int_0^1 du\,\e^{\rho^2 v^2(1+u^2)}
\left\{J_0(v u x) J_0(v y)- J_0(v x) J_0(v u y)\right\}
,\nonumber\\
&&I(x,y)=
\frac{ \sqrt{x y}}{2}  \int_\pi^\infty dv\, v^2 \,\e^{-2 \rho^2 v^2} 
\left\{x J_1(v x) J_0(v y)- J_0(v x)y J_1(v y)\right\}
\label{chGSEchGUE}
\end{eqnarray}
}%
for chGSE-chGUE crossover.
Here $x_i\equiv\lambda_i/\varDelta(0)$ 
are unfolded eigenvalues.
The probability $E(s)$ that the interval $[0, s]$ contains no eigenvalues is given as the Fredholm determinant
\be
E(s)=
{\rm Det} (\openone-\hat{K}_s)^{1/2},
\label{FredholmPfaffian}
\ee
where $\hat{K}_s$ is an integral operator of convolution with the dynamical Bessel kernel $K(x,y)$ 
(\ref{Pfaffian})-(\ref{chGSEchGUE})
acting on two-component $L^2$-functions $f$ over the interval $[0,s]$,
$(\hat{K}_s f)(x)=\int_0^s dy\,K(x,y)f(y)$.
The probability distribution $p_1(s)$ of the unfolded smallest eigenvalue $s=\lambda_1/\varDelta(0)$,
which is half of the very central level spacing, 
is given by the first derivative $p_1(s)=-E'(s)$.
See Ref.\citen{damn} for an alternative derivation.

In the limit
$x, y\gg 1$ with $r=x-y$ kept finite,
the eigenvalues $x, y$ are liberated from repulsion by their mirror images $-x, -y$ 
and their correlation becomes translationally invariant.
The $S, D, I$ components of the kernel 
in this limit are given
\nolinebreak 
by,\cite{meh}
\bea
&&
S(r)=\frac{\sin \pi r}{\pi r},\ 
D(r)=\int_0^\pi \frac{dv}{\pi}v\,\e^{2\rho^2 v^2}\sin v r,\ 
I(r)=\int_\pi^\infty \frac{dv}{\pi v}\e^{-2\rho^2 v^2}\sin  v r, 
\label{GOEGUE}\\
&&
S(r)=\frac{\sin \pi r}{\pi r},\ 
D(r)=\int_\pi^\infty \frac{dv}{\pi}v\,\e^{-2\rho^2 v^2}\sin v r,\ 
I(r)=\int_0^\pi \frac{dv}{\pi v}\e^{2\rho^2 v^2}\sin  v r,
\label{GSEGUE}
\eea
for GOE-GUE and GSE-GUE crossover, respectively.
These expressions can be used to interpolate between two nonchiral ensembles,
GOE (GSE) at $\rho=0$ and GUE at $\rho=\infty$.
The gap probability $E(s)$
is again given as (\ref{FredholmPfaffian}) with this dynamical sine kernel\cite{mp}.
The distribution $P(s)$ of the level spacings $s=x_{i+1}-x_i$
is given by the second derivative $P(s)=E''(s)$.

An efficient way of evaluating the Fredholm determinant of
a trace-class 
operator $\hat{K}_s$ acting on $L^2$-functions over an interval $[0,s]$
is the Nystr\"{o}m-type discretization\cite{bor}
\be
{\rm Det}(1-\hat{K}_s)\simeq \det \left[\delta_{ij}-K(x_i,x_j) \sqrt{w_i\,w_j}\right]_{i,j=1}^m~.
\label{Nystrom}
\ee
Here, the quadrature rule consists of a set of points $\{x_i\}$ 
taken from the interval $[0,s]$ and associated weights $\{w_i\}$ such that
${\int_0^s f(x)dx  \simeq \sum_{i=1}^m f(x_i) w_i}$.
As the order $m$ of the approximation increases, 
the RHS of (\ref{Nystrom}) is proven to converge uniformly to its LHS.
The convergence is rapid and exponentially fast \cite{bor}.
For our purpose, we employ the Gauss quadrature rule (sampling at the Legendre nodes), as 
15-digit accuracy is known to be attainable already with only $m=5$
for the Fredholm determinant $E(0.1)$ for the sine kernel\cite{bor}.
We have applied the Nystr\"{o}m-type method 
to the dynamical Bessel kernel (\ref{Pfaffian}), (\ref{chGOEchGUE}), (\ref{chGSEchGUE})
and the dynamical sine kernel (\ref{Pfaffian}), (\ref{GOEGUE}), (\ref{GSEGUE}),
and evaluated $p_1(s)$ and $P(s)$ for chG(O,S)E-chGUE and G(O,S)E-GUE crossover, respectively.
In order to achieve the accuracy needed for computing the first and second derivatives
($p_1(s)$ and $P(s)$) to a high precision, we chose the approximation order $m$ to be 
at least 20 for the former and 100 for the latter, 
and confirmed the stability of the results for increasing $m$.
Plots of $p_1(s)$ and $P(s)$
for the region $0\leq s\leq 3-4$ and for the parameter range $\rho\lesssim 1$
are exhibited 
in Fig.~1.
\begin{figure}[t]
\begin{center}
\includegraphics[bb=0 0 202 129]{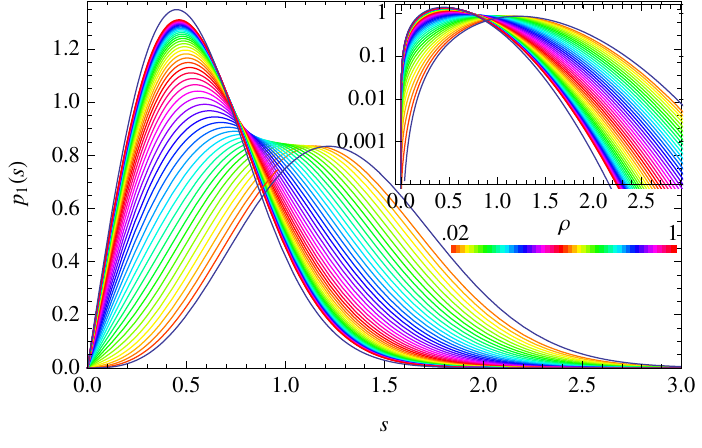}\includegraphics[bb=0 0 202 129]{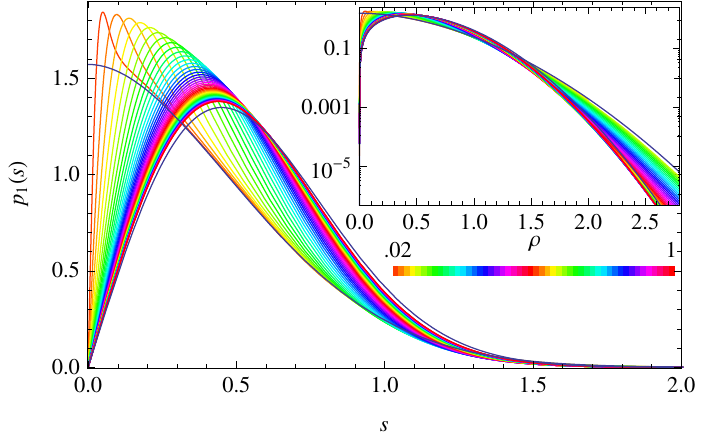}\\
\includegraphics[bb=0 0 202 131]{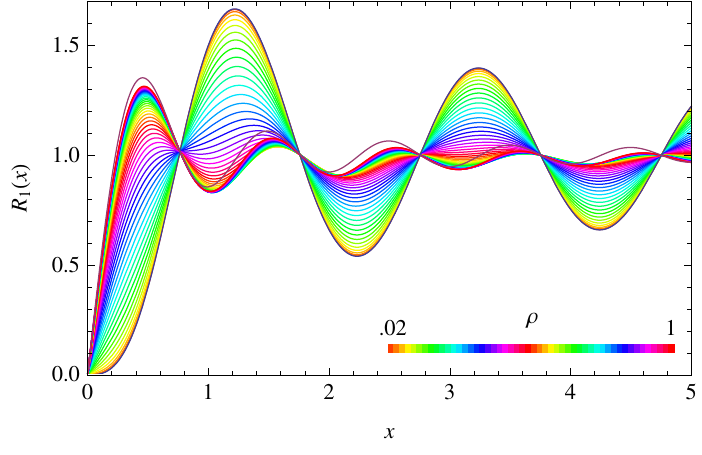}\includegraphics[bb=0 0 202 129]{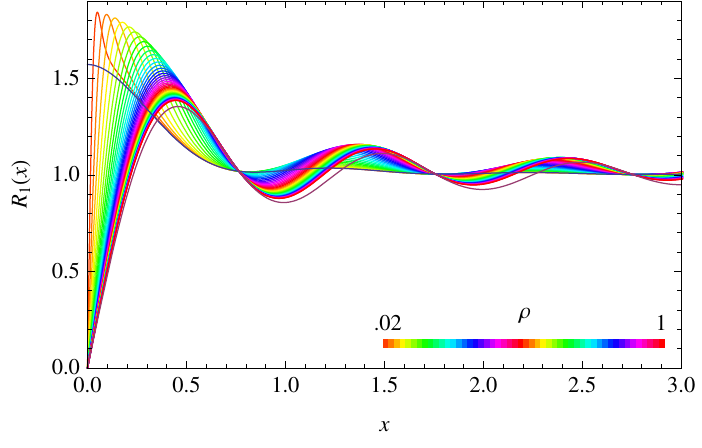}\\
\includegraphics[bb=0 0 202 136]{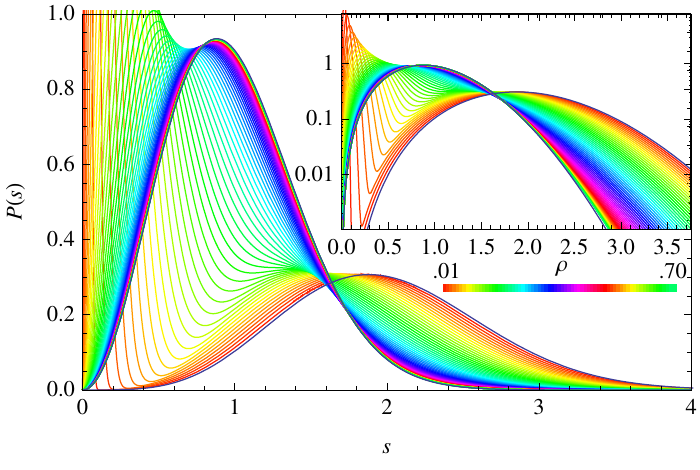}\includegraphics[bb=0 0 202 134]{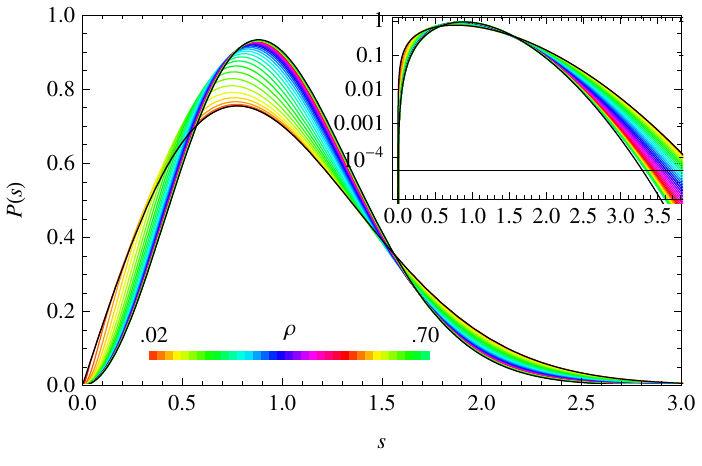}
\caption{
Smallest eigenvalue distributions (top),
microscopic level densities (center) and
level spacing distributions (bottom)
for (ch)GSE-(ch)GUE (left) and (ch)GOE-(ch)GUE (right) crossover.}
\end{center}
\vspace{-4mm}
\end{figure}
Although Mehta and Pandey\cite{mp} have expressed $P(s)$
in terms of eigenvalues of an infinite-dimensional matrix,
the elements of which are integrals involving prolate spheroidal functions,
numerical plots of $p_1(s)$ for 
parametric chiral RM ensembles have never appeared in the literature.
We also exhibit single-level densities $R_1(x)$ (\ref{Pfaffian}) in Fig.~1,
whose first peaks comprise $p_1(s)$.
The practical advantage of adopting distributions of individual level spacings
over $n$-level correlation functions ($R_1(x)$, etc.) for fitting is clear from the figures.
As the oscillation of the latter consists of overlapping multiple peaks,   
the characteristic shape of each peak is inevitably smeared, 
to yield a rather structureless curve for which an accurate fit is difficult.
On the other hand, the shape of the former is clearly distinguishable and
is extremely sensitive to the $\rho$ parameter, because
the ratio of $p_1(s)$ or $P(s)$ for the orthogonal and symplectic classes to that for the unitary class
grows as $\exp \frac{\pi^2s^2}{16}$ for large $s$. 
Therefore, $p_1(s)$ and $P(s)$ admit very sharp one-parameter fitting
by the least-squares method or (in principle) simply from the tails of
the curves.

\vspace*{1mm}
\noindent
{\it Dirac spectrum}\hspace*{7mm}
The Dirac operator of QCD-like theory with quarks in a real or pseudoreal representation
possesses antiunitary symmetry unlike those in complex representations \cite{ver}: 
the Euclidean Dirac operator for quarks in the fundamental (adjoint) representation of $SU$(2) commutes with $C\tau_{2}K$ ($CK$).
Here, $C$ is the charge conjugation matrix,
$\tau_2$ is one of the generators of the gauge group
and $K$ is the complex conjugation.
As $(C\tau_{2}K)^2=+1$ ($(CK)^2=-1$), $D$ is essentially a real symmetric
(quaternion self-dual) matrix.
The reality and self-duality of the continuum Dirac operators are known to be
interchanged for the lattice staggered Dirac operators
owing to the absence of the charge conjugation matrix \cite{hv}.
Since the $U(1)$ Dirac operator
in a continuum or on a lattice  
possesses no such antiunitary symmetry,
the Dirac operator in the fundamental representation of $SU$(2) (the adjoint representation of $SU$($N$))$\times U(1)$
has its antiunitary symmetry (weakly) broken.
As the simplest example,
we consider $SU$(2) quenched lattice gauge theory
under the twisted boundary condition,
that is, we multiply $SU$(2) link variables on the temporal boundary 
of the hypercubic lattice of size $V=L^4$ by a constant phase $\e^{i\theta_{n,\mu}}$ with
$\theta_{n,\mu}= 2\pi  \varphi\, \delta_{n_4,L}\delta_{\mu,4}$
$(\varphi\ll1)$.
As $SU$(2) Dirac operators possess (pseudo)reality either for the periodic or antiperiodic boundary condition
on each dimension,
we impose periodicity (antiperiodicity) for the spatial (temporal) direction
and consider a small deviation (twisting) in the temporal boundary condition.
This twisting is gauge-equivalent to 
a fixed $U(1)$ background 
of flux $2\pi\varphi$ 
or imaginary chemical potential $\mu= {i 2\pi \varphi}/{L}$, 
which is the measure of symmetry violation 
and plays the r\^{o}le of 
$\alpha$ in the parametric RMs\cite{dm}. 
Its effect on the chiral Lagrangian is completely dictated\cite{svd},
in parallel with the case of real chemical potentials \cite{kogut}.
 
For our aim of confirming the presence of chG(S,O)E-chGUE crossover
in lattice gauge theories,
we restrict ourselves to the strong coupling region of $SU$(2) at $\beta=4/g^{2}=0-1$, 
where the level density at the origin, $1/\varDelta(0)$, is sufficiently above zero 
and the chiral symmetry is spontaneously broken.
Accepting that the model is away from the continuum limit,
we employ the simplest algorithm:  
unimproved plaquette action and
a 10-hit heat-bath update coupled with overrelaxation.
Because of our need to detect possibly small deviations of spectral fluctuation
from the universal RM statistics at either end ($\rho=0$ or $\infty$),
we give priority to the number of independent gauge configurations 
and perform our simulation on a lattice of the smallest size $V=4^4$.
This choice is sufficient for  
measuring the 
local
behavior of eigenvalues within $3\varDelta-4\varDelta$
and determining the $\rho$ parameter precisely.
In this region, the systematic deviation due to the smallness
of the lattice is 
expected to be less prominent
(it will manifest itself at larger separation) than the statistical fluctuation.

Dirac spectral statistics are fitted to parametric RM predictions by the following steps:
(i) First we perform pure $SU$(2) simulations for each $\beta$ and measure the smallest
fundamental or adjoint staggered Dirac eigenvalue $\lambda_1$ for $O(10^5)$ configurations.
Taking for granted that Dirac eigenvalues obey chG(S,O)E statistics, 
we find the value of $\varDelta$ that optimally
fits the histogram of the smallest Dirac eigenvalue
to the rescaled chG(S,O)E result $p_1(\lambda_1/\varDelta)/\varDelta$.
(ii) Next, we multiply the $SU$(2) link variables by the twisting phases and
measure all Dirac eigenvalues $\{\lambda_i\}$
for $N_{{\rm conf}}=O(10^{4})$ independent configurations.
The unfolded smallest eigenvalue is still defined by $x_1=\lambda_1/\varDelta(0)$ with
respect to $\varDelta(0)$ determined from pure $SU$(2) simulations. 
We fit the frequencies of $x_1$ to the RM prediction $p_1^{(\rho)}(s)$ by the least-squares method
and find the optimal value of $\rho$.
The valid range of fitting the smallest eigenvalue $x_1$ is chosen
to be [0, 2.8] (fundamental) and [0, 1.6] (adjoint), divided into 20 segments.
(iii) Finally, we measure the distribution of level spacings from the spectral `plateau' 
$[\lambda_{\rm m},\lambda_{\rm M}]$ 
adjacent to but not including the origin,
in which the mean level spacing $\varDelta(\lambda)$
is well approximated as a constant close to $\varDelta(0)$.
In order to avoid possible distortion of the level spacing distribution, 
we take $\lambda_{\rm m}$ not too close to the origin
and set its smallest value to be the $11^{\rm th}$ eigenvalue.
We fit the frequencies of unfolded level spacings
$s=(\lambda_{i+1}-\lambda_i)/\varDelta(\bar{\lambda}),\ \lambda_i, \lambda_{i+1}\in[\lambda_{\rm m},\lambda_{\rm M}]$
to the RM prediction $P^{(\rho)}(s)$ as before.
Thanks to the enormous gain in statistics resulting from the spectral averaging,
we can safely set the fitting range to be as large as [0, 3.8] and divide it into
40 segments.

\vspace*{1mm}
\noindent
{\it Simulation results}\hspace*{7mm} 
We generated 40000 independent configurations
at each value of the coupling constant $\beta=0, 0.5, 1$
and the twisting $\varphi=0.01-.06$
on a lattice of dimensions $V=4^4$.
Optimal values of $\rho$ determined from the
smallest eigenvalue distributions (SED) and level spacing distributions (LSD)
for the fundamental (F) and adjoint (A) representations
are tabulated in the central columns of Table I.
\begin{figure}[htb]
\begin{center}
\includegraphics[bb=0 0 260 169,scale=0.75]{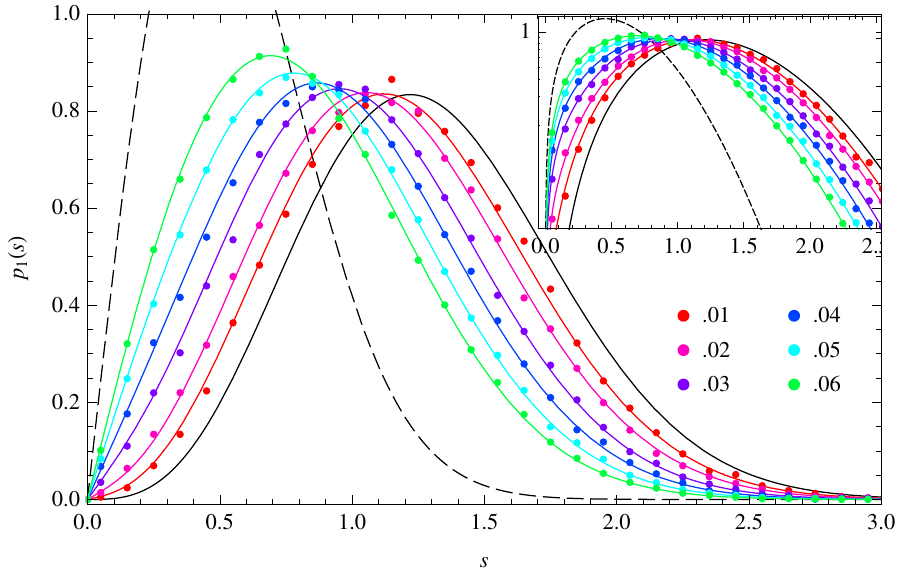}\includegraphics[bb=0 0 260 165,scale=0.75]{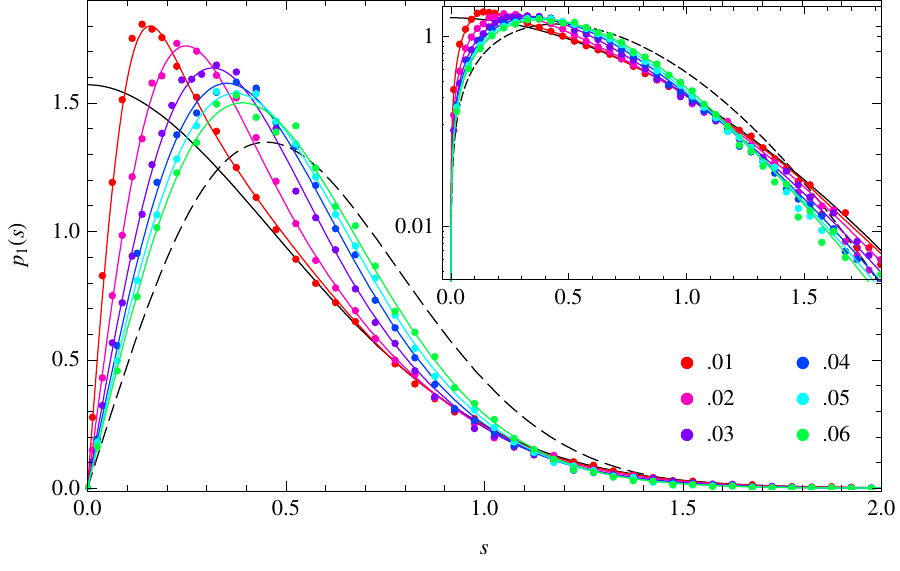}\\
\includegraphics[bb=0 0 260 167,scale=0.75]{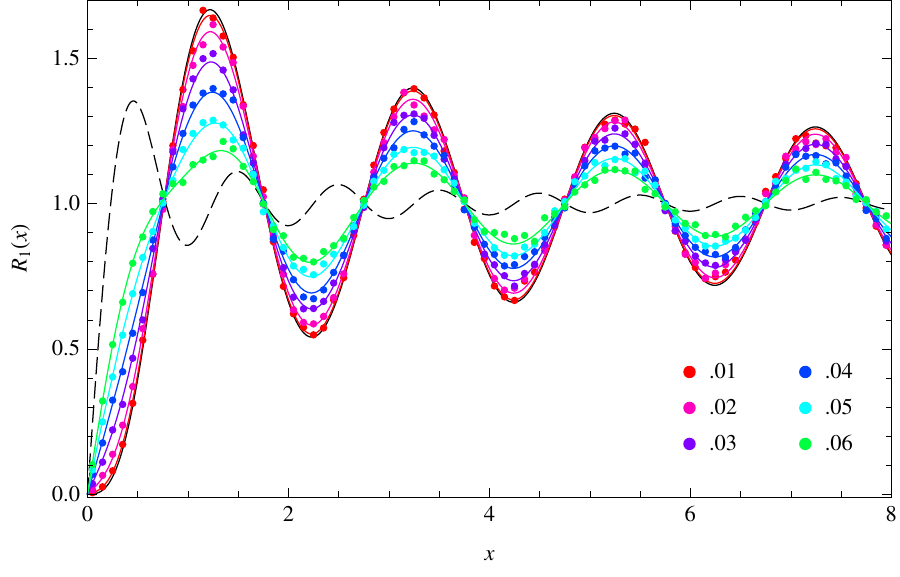}\includegraphics[bb=0 0 260 165,scale=0.75]{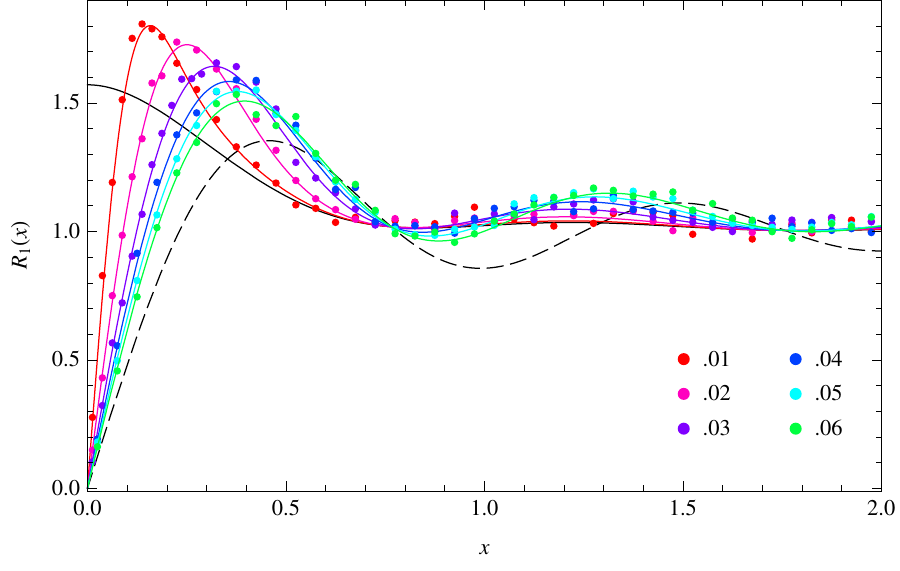}\\
\includegraphics[bb=0 0 260 168,scale=0.75]{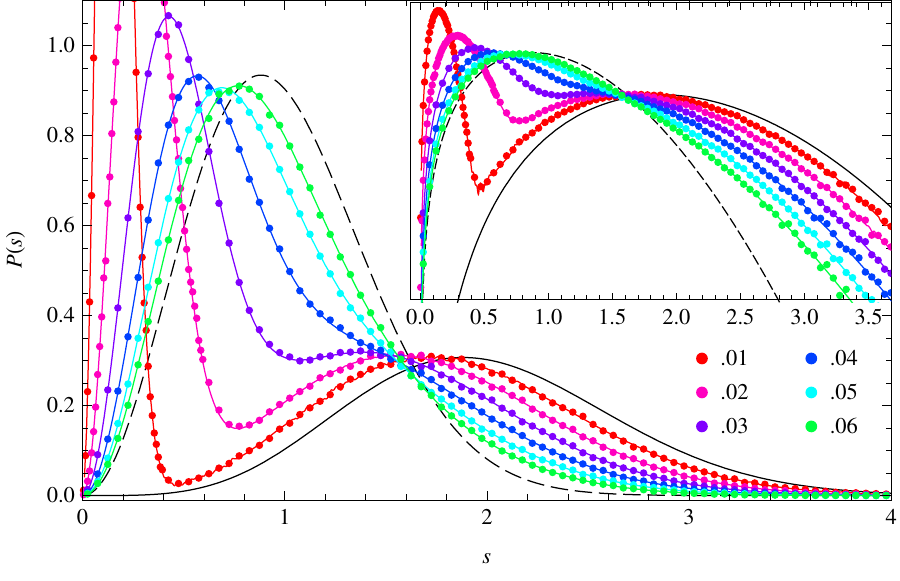}\includegraphics[bb=0 0 260 170,scale=0.75]{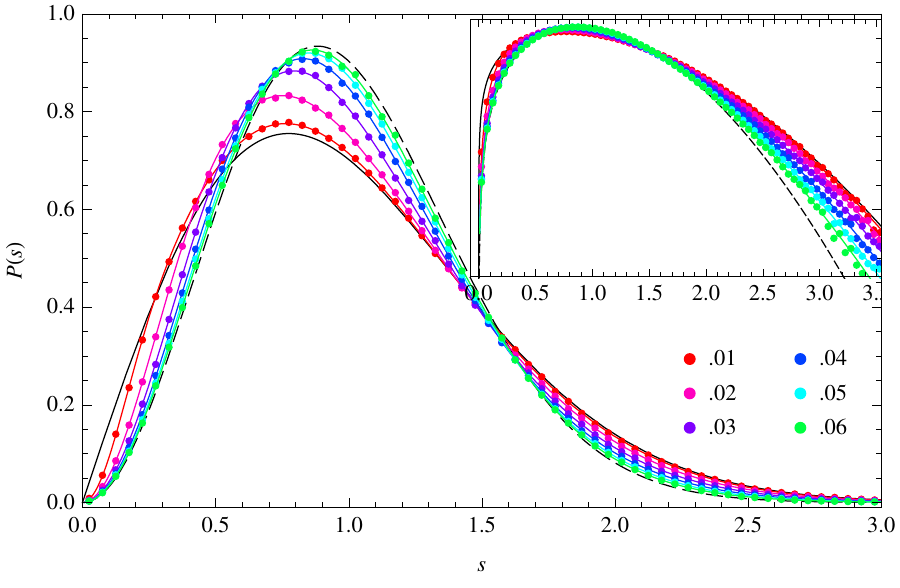}
\caption{
Smallest eigenvalue distributions (top),
microscopic level densities (center) and
level spacing distributions (bottom)
of $SU$(2) fundamental (left) and adjoint (right) staggered Dirac operators
at coupling $\beta=0.5$ and flux $\varphi=0.01-.06$. $V=4^4$, $N_{\rm conf}=40000$. 
}
\end{center}
\end{figure}
\begin{table}[htb]
\centering
\caption{Crossover parameters and low-energy constants.}\label{table1}
\setlength{\tabcolsep}{1.3mm}
\begin{tabular}{|c|c||c|c|c|c|c|c|c|c|c|c|c||}
\hline
\hline
$\beta$&\!\!\!rep/dist\!\!\!&$\varDelta$&$\Sigma$&
\multicolumn{6}{|c|}{$\rho$}
&$\underline{\sqrt{\varDelta}\rho}$&$\underline{\left.F^2\right.}$&$F^2$ \\ \cline{5-10}
&&&&$\varphi$=.01&.02&.03&.04&.05&.06&$\mu$&$\Sigma$&\\
\hline
\hline
\!\!\!0  \!\!\!&\!\!\!\! F/SED \!\!\!\!&\!\!\!\! .00930(2) \!\!\!\!&\!\!\!\! 1.319(2) \!\!\!\!&\!\!\!\! .059(1) \!\!\!\!&\!\!\!\! .121(1) \!\!\!\!&\!\!\!\! .182(1) \!\!\!\!&\!\!\!\! .239(2) \!\!\!\!&\!\!\!\! .302(2) \!\!\!\!&\!\!\!\! .363(2) \!\!\!\!&\!\!\!\! .370(1) \!\!\!\!&\!\!\!\! .215(1) \!\!\!\!&\!\!\!\! -\!\!\!\!    \\ 
\!\!\!   \!\!\!&\!\!\!\! F/LSD \!\!\!\!&\!\!\!\! .00929(0) \!\!\!\!&\!\!\!\! -         \!\!\!\!&\!\!\!\! .061(0) \!\!\!\!&\!\!\!\! .122(0) \!\!\!\!&\!\!\!\! .182(1) \!\!\!\!&\!\!\!\! .244(1) \!\!\!\!&\!\!\!\! .304(2) \!\!\!\!&\!\!\!\! .364(4) \!\!\!\!&\!\!\!\! .374(1) \!\!\!\!&\!\!\!\! .220(1) \!\!\!\!&\!\!\!\! .290(1)\!\!\!\! \\ \cline{2-13}
\!\!\!   \!\!\!&\!\!\!\! A/SED \!\!\!\!&\!\!\!\! .00620(1) \!\!\!\!&\!\!\!\! 1.980(3) \!\!\!\!&\!\!\!\! .076(1) \!\!\!\!&\!\!\!\! .150(2) \!\!\!\!&\!\!\!\! .224(3) \!\!\!\!&\!\!\!\! .288(5) \!\!\!\!&\!\!\!\! .373(7) \!\!\!\!&\!\!\!\! .435(9) \!\!\!\!&\!\!\!\!  .372(3) \!\!\!\!&\!\!\!\! .217(3) \!\!\!\!&\!\!\!\! -  \!\!\!\!  \\ 
\!\!\!   \!\!\!&\!\!\!\! A/LSD \!\!\!\!&\!\!\!\! .00618(0) \!\!\!\!&\!\!\!\! -         \!\!\!\!&\!\!\!\! .075(3) \!\!\!\!&\!\!\!\! .149(4) \!\!\!\!&\!\!\!\! .223(6) \!\!\!\!&\!\!\!\! .298(9) \!\!\!\!&\!\!\!\! .371(14)\!\!&\!\!\!\! .444(22)\!\!&\!\!\!\! .372(5) \!\!\!\!&\!\!\!\! .218(6) \!\!\!\!&\!\!\!\! .432(11)\! \!\!\!\!\\ \cline{2-13}
\hline
\!\!\!0.5\!\!\!&\!\!\!\! F/SED \!\!\!\!&\!\!\!\! .01014(1) \!\!\!\!&\!\!\!\! 1.210(1) \!\!\!\!&\!\!\!\! .052(1) \!\!\!\!&\!\!\!\! .102(1) \!\!\!\!&\!\!\!\! .156(1) \!\!\!\!&\!\!\!\! .206(2) \!\!\!\!&\!\!\!\! .257(1) \!\!\!\!&\!\!\!\! .308(2) \!\!\!\!&\!\!\!\!  .330(1) \!\!\!\!&\!\!\!\! .171(1) \!\!\!\!&\!\!\!\! -   \!\!\!\! \\
\!\!\!   \!\!\!&\!\!\!\! F/LSD \!\!\!\!&\!\!\!\! .01004(0) \!\!\!\!&\!\!\!\! -         \!\!\!\!&\!\!\!\! .052(0) \!\!\!\!&\!\!\!\! .103(0) \!\!\!\!&\!\!\!\! .155(1) \!\!\!\!&\!\!\!\! .206(1) \!\!\!\!&\!\!\!\! .257(1) \!\!\!\!&\!\!\!\! .307(2) \!\!\!\!&\!\!\!\! .329(1) \!\!\!\!&\!\!\!\! .170(1) \!\!\!\!&\!\!\!\! .208(1)\!\!\!\! \\ \cline{2-13}
\!\!\!   \!\!\!&\!\!\!\! A/SED \!\!\!\!&\!\!\!\! .00629(1) \!\!\!\!&\!\!\!\! 1.951(2) \!\!\!\!&\!\!\!\! .066(1) \!\!\!\!&\!\!\!\! .131(2) \!\!\!\!&\!\!\!\! .197(3) \!\!\!\!&\!\!\!\! .258(4) \!\!\!\!&\!\!\!\! .312(5) \!\!\!\!&\!\!\!\! .370(7) \!\!\!\!&\!\!\!\! .324(2)  \!\!\!\!&\!\!\!\! .165(2) \!\!\!\!&\!\!\!\! -   \!\!\!\! \\
\!\!\!   \!\!\!&\!\!\!\! A/LSD \!\!\!\!&\!\!\!\! .00622(0) \!\!\!\!&\!\!\!\! -         \!\!\!\!&\!\!\!\! .067(3) \!\!\!\!&\!\!\!\! .134(4) \!\!\!\!&\!\!\!\! .202(5) \!\!\!\!&\!\!\!\! .268(7) \!\!\!\!&\!\!\!\! .331(9) \!\!\!\!&\!\!\!\! .400(18)\!\!\!\!&\!\!\!\! .336(4) \!\!\!\!&\!\!\!\! .177(4) \!\!\!\!&\!\!\!\! .350(9)\!\!\!\! \\ \cline{2-13}
\hline
\!\!\! 1 \!\!\!&\!\!\!\! F/SED \!\!\!\!&\!\!\!\! .01132(13)\!\!\!&\!\!\!\! 1.084(12)\!\!\!&\!\!\!\! .046(2) \!\!\!\!&\!\!\!\! .094(1) \!\!\!\!&\!\!\!\! .138(1) \!\!\!\!&\!\!\!\! .185(1) \!\!\!\!&\!\!\!\! .233(2) \!\!\!\!&\!\!\!\! .280(2) \!\!\!\!&\!\!\!\! .315(1)  \!\!\!\!&\!\!\!\! .156(1) \!\!\!\!&\!\!\!\! - \!\!\!\!   \\
\!\!\!  \!\!\!&\!\!\!\! F/LSD \!\!\!\!&\!\!\!\! .01105(0) \!\!\!\!&\!\!\!\! -         \!\!\!\!&\!\!\!\! .048(0) \!\!\!\!&\!\!\!\! .095(0) \!\!\!\!&\!\!\!\! .143(1) \!\!\!\!&\!\!\!\! .190(1) \!\!\!\!&\!\!\!\! .237(1) \!\!\!\!&\!\!\!\! .284(2) \!\!\!\!&\!\!\!\! .319(1) \!\!\!\!&\!\!\!\! .160(1) \!\!\!\!&\!\!\!\! .177(1) \!\!\!\!\\ \cline{2-13}
\!\!\!  \!\!\!&\!\!\!\! A/SED \!\!\!\!&\!\!\!\! .00633(2) \!\!\!\!&\!\!\!\! 1.939(7) \!\!\!\!&\!\!\!\! .066(2) \!\!\!\!&\!\!\!\! .134(2) \!\!\!\!&\!\!\!\! .198(3) \!\!\!\!&\!\!\!\! .257(4) \!\!\!\!&\!\!\!\! .320(6) \!\!\!\!&\!\!\!\! .391(7) \!\!\!\!&\!\!\!\! .332(2) \!\!\!\!&\!\!\!\! .173(2) \!\!\!\!&\!\!\!\! -  \!\!\!\! \\ 
\!\!\!  \!\!\!&\!\!\!\! A/LSD \!\!\!\!&\!\!\!\! .00632(0) \!\!\!\!&\!\!\!\! -         \!\!\!\!&\!\!\!\! .067(4) \!\!\!\!&\!\!\!\! .133(4) \!\!\!\!&\!\!\!\! .198(5) \!\!\!\!&\!\!\!\! .265(7) \!\!\!\!&\!\!\!\! .331(11)\!\!\!\!&\!\!\!\! .398(16)\!\!\!\!&\!\!\!\! .335(4) \!\!\!\!&\!\!\!\! .176(5) \!\!\!\!&\!\!\!\! .342(9) \!\!\!\!\\ \cline{2-13}
\hline
\hline
\end{tabular}
\end{table}
\noindent
Sample plots of $p_1(s)$, $R_1(x)$ and $P(s)$ at $\beta=0.5$ are
exhibited in Fig.~2.
In all figures, measured histograms are plotted by colored dots, and optimally fit 
parametric RM results are shown as curves of the same color.
Also plotted in the figures are the results from chG(S,O)E or G(S,O)E (black real lines),
and chGUE or GUE (broken lines).
Note that the microscopic level densities $R_1(x)$ are not fitted to the corresponding data,
but are merely obtained by substituting the values of $\rho$ determined from the SEDs.
Even by inspection, the precision of one-parameter fitting is convincing in all distributions shown.
The ranges of $\chi^2$/d.o.f.\ for the optimal RM distributions are
$0.53-1.47$ (Fund/SED),
$0.56-1.57$ (Adj/SED),
$0.64-1.49$\footnote{Except for Fund/LSD at very small $\rho\lesssim 0.1$,
where the distribution becomes extremely peaky at small $s$ owing to the onset of
Kramers degeneracy and the fitting error is inevitably enhanced.} (Fund/LSD),
$0.68-1.30$ (Adj/LSD).
Considering the smallness of our lattice, the precision achieved is astonishing;
it is even more surprising when one recalls the acute sensitivity of $P^{(\rho)}(s)$ to $\rho$ shown in Fig.~1.
The goodness of fit is
comparable to those in the pioneering papers \cite{dam1},
which involved fitting to the spectral data from larger lattices: 
$\chi^2$/d.o.f.$=0.33$ for quenched QCD on a $12^4$ lattice and
$\chi^2$/d.o.f.$=1.13-1.33$ for dynamical QCD on a $6^4$ lattice, with
$O(10^3)$ configurations. 
\begin{wrapfigure}{r}{6.6cm}
\centerline{\includegraphics[bb=0 0 191 120]{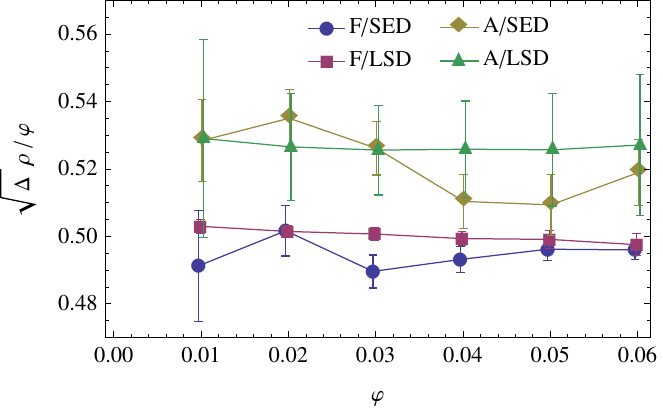}}
\caption{
Ratios between the twisting $\varphi$ and the
crossover parameter $\sqrt{\varDelta}\rho$ at $\beta=1$.
}
\end{wrapfigure}
From Table I one notices that (i)
for a fixed gauge coupling $\beta$, the relationships between $\varphi$ and $\rho$ are all reasonably linear. 
In Fig.~3, we present a sample plot of
${\sqrt{\varDelta}}\rho/\varphi$ at $\beta=1$.
The error bars indicated are statistical only.
The mean values of this ratio, namely from LSDs, are very stable under the change of flux $\varphi$
and fluctuate only within 0.5\%,
and
(ii) the values of $\rho$ determined from SED and from LSD are in
good agreement, as they should be. High
linearity 
will be 
essential for the precise determination of the pion decay constant.

\vspace*{1mm}
\noindent
{\it Low-energy constants}\hspace*{7mm} 
The effective low-energy Lagrangian for QCD-like theories with $N_F$ flavors of quarks
in a (pseudo)real representation
at a finite chemical potential $\mu$ and bare quark mass $m$
is unambiguously fixed by the global symmetry alone 
(as long as $\mu$ is much smaller than the vector meson mass)
and takes a form containing two phenomenological constants:
$F$ the `pion' decay constant and $\Sigma=\<\bar{\psi}\psi\>/N_F$ the chiral condensate,
both measured in the chiral and zero-chemical potential limit $m,\, \mu\to 0$.
If the theory is in a finite volume $V=L^4$ and the
Thouless energy defined as
${E_c\sim {F^2}/{\Sigma L^2}}$
is much larger than $m$, 
the path integral is dominated by the zero mode and takes the tractable form
\be
Z=\int_{{SU}(2N_F)} \!\!\!\!\!\!\!\!\!\!\!\!\! dU\,\,\,\exp\Bigl(
V\mu^2 F^2\,{\rm tr}\, (\hat{B} U^\dagger\hat{B} U+\hat{B}\hat{B})
+\frac12
 V\Sigma m \,{\rm Re}\,{\rm tr}\, \hat{M}U
\Bigr).
\label{Zchiral}
\ee
Here, $U$ is an $SU$($2N_F$) matrix,
$\hat{B}=\sigma_3 \otimes \openone_{N_F}$ and
$\hat{M}=\sigma_1 \otimes \openone_{N_F}\ \(i\sigma_2 \otimes \openone_{N_F}\)$ 
for quarks in a real (pseudoreal) representation.
In order to extract the Dirac spectrum,
one introduces fermionic as well as bosonic quarks in the fundamental theory,
leading to the graded group version of (\ref{Zchiral}) on the effective theory side.
Parametrizing the graded matrix $U$ in terms of its eigenvalues and
comparing the resulting expression
(after analytic continuation $\mu\to i\mu$ and $m\to i\lambda$)
with the RM results (\ref{chGOEchGUE}) and (\ref{chGSEchGUE}),
the coefficients of the chemical-potential and `mass' terms in the exponents on both sides
are readily identified as
$4VF^2\mu^2=2\pi^2\rho^2$ and $V\Sigma \lambda=\pi x$, respectively.
Using the definition of the unfolded eigenvalues $x=\lambda/\varDelta$,
the latter gives the Banks-Casher relation $\Sigma=\pi/\varDelta V$,
which determines one of the low-energy constants $\Sigma$ in terms of $\varDelta$.
Eliminating the volume in favor of the level spacing, the former equation becomes
\pagebreak
\be
{\sqrt{\varDelta}}\,\rho=\sqrt{\frac{2 }{\pi}\frac{F^2}{\Sigma }} \mu
=\sqrt{\frac{2 }{\pi}\frac{F^2}{\Sigma }} \frac{2\pi}{L}\varphi,
\ee
where the left-hand side is a volume-independent combination.
Accordingly, one can determine another low-energy constant
$F^2/\Sigma$ from the slope of
$\varphi$-${\sqrt{\varDelta}}\rho$ plots, preferably on lattices of various sizes.
In the next-to-rightmost column of Table I we exhibit the values of $F^2/\Sigma$,
with all the numerals in the lattice unit.

\begin{wrapfigure}{1}{6.6cm}
\hspace*{-1.5mm}
\includegraphics[bb=0 0 94 116]{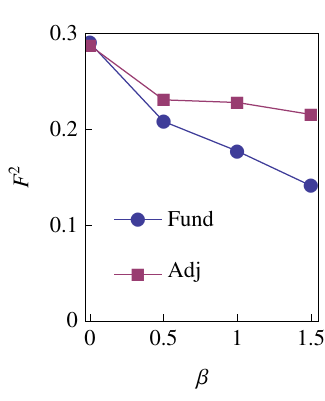}
\includegraphics[bb=0 0 94 116]{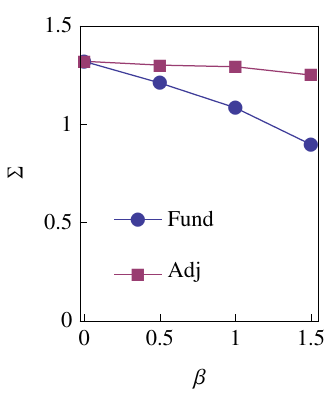}
\caption{
Low-energy constants $F^2$ and $\Sigma$
for $SU$(2) quenched lattice gauge theory.
}
\end{wrapfigure}
Note that in the parameter region $V\Sigma |m|\gg1$, Eq.~(\ref{Zchiral}) should
approach the $\sigma$ model of nonchiral parametric RM ensembles \cite{aie}, 
but the pion decay or diffusion constant multiplying ${\rm tr}\, \hat{B} U^\dagger\hat{B}U$ is unaffected.
Accordingly, if the mean level spacing is approximately constant in a window in the very vicinity of the origin,
one can determine $F^2/\Sigma$ from the {\it bulk} correlation (namely the LSD)
in that window. 
As LSDs admit the high-precision determination of $\rho$ and
achieve much higher linearity of the $\varphi$-${\sqrt{\varDelta}}\rho$ plots
than those from SEDs (Fig.~3), 
we have adopted the former for the determination of $F^2/\Sigma$.
The coupling dependence of these low-energy constants is 
summarized in Table I 
and plotted in Fig.~4
(including the result from simulations at $\beta=1.5$).
In order to offset the number of components in the gauge multiplet,
the plots of the adjoint are multiplied by 2/3.
At $\beta=0$, both low-energy constants agree between the fundamental and adjoint representations.
This observation is consistent with the fact 
that in the strong coupling limit 
these constants are common to the large-$N$ $Sp$($2N$) and $O(N$) lattice gauge theories \cite{nn},
which share the same antiunitary symmetries as the $SU$(2) fundamental and adjoint.

\vspace*{1mm}
\noindent
{\it Summary}\hspace*{7mm} 
We have evaluated the smallest eigenvalue and level spacing distributions
for (ch)GSE-(ch)GUE and (ch)GOE-(ch)GUE crossover
using a Nystr\"{o}m-type method, the former being our new contribution.
These RM results are applied to fit
the fundamental and adjoint staggered Dirac spectra of 
$SU$(2) quenched lattice gauge theory
under the twisted boundary condition.
Excellent one-parameter fitting 
is achieved for all cases of interest.
The acute sensitivity of our fitting distributions, $p_1(s)$ and $P(s)$, 
on the crossover parameter $\rho$
leads to the precise determination of the pion decay constant $F$
from its twisting dependence.
This method, feasible on a small-size lattice, 
has a clear advantage over the conventional method
using axial correlators, which inevitably requires a large temporal dimension.

Our treatment is complementary to the previous approach of determining $F$ for
two-color QCD from its Dirac spectrum \cite{ake}, which conversely measured
the response of {\it complex} Dirac eigenvalues to a {\it real} chemical potential.
From a practical point of view,
the use of imaginary chemical potential is advantageous because
(i) it does not require the projection of eigenvalues to the real or imaginary axis
for fitting, which is usually requisite for dealing with complex eigenvalues, and 
(ii) the two-dimensional motion of complex eigenvalues may lead to large statistical fluctuation.
Similar complementary treatments were applied for
three-color QCD at real and imaginary {\em isospin} chemical potentials,
corresponding to non-Hermitian chiral RMs \cite{dam1}
and Hermitian crossover chiral RMs \cite{dam2,ai}, respectively.
Combined with the results reported here, which fill the vacancy, 
the established fact that Dirac spectra in all three cases 
($SU$(2)-fund.+$\mu$, $SU$(2)-adj.+$\mu$, $SU$(3)-fund.+$\mu_{\rm iso}$)
agree perfectly with
predictions from corresponding zero-mode-approximated chiral Lagrangians
in both regions of $\mu^2\gtrless0$
constitutes encouraging evidence 
for the validity of analytic continuation in the $\mu$-plane.

Finally, we note that our preliminary simulation has affirmed that the Dirac spectra of
$SU$(2)$\times U(1)$ quenched lattice gauge theory also fit excellently to the parametric RM predictions. 
We are currently accumulating 
data on 
lattices of larger size than that treated here, 
and the results will be reported in a subsequent publication.

I thank T. Nagao and A. Nakamura for valuable communications and discussions.
Discussions during the workshop ``Field Theory and String Theory'' 
(YITP-W-12-05) 
at YITP, Kyoto University, were useful in the final stage of this work.

\end{document}